\documentclass{article}

\pdfoutput=1
\usepackage[nonatbib]{nips_2016}

\usepackage{nips_2016}

\usepackage[utf8]{inputenc} % allow utf-8 input
\usepackage[T1]{fontenc}    % use 8-bit T1 fonts
\usepackage{hyperref}       % hyperlinks
\usepackage{url}            % simple URL typesetting
\usepackage{booktabs}       % professional-quality tables
\usepackage{amsfonts}       % blackboard math symbols
\usepackage{nicefrac}       % compact symbols for 1/2, etc.
\usepackage{microtype}      % microtypography

\usepackage{algorithmic}    % for pseudo code
\usepackage{listings}       % listings

\usepackage{adjustbox}
\usepackage{csquotes}

\author{
  Edward J. Yoon \\
  Samsung Electronics\\
  56, Seongchon-gil, Seocho-gu, Seoul, Korea \\
  \texttt{edward.yoon@samsung.com} \\
}

\usepackage[utf8]{inputenc}
\usepackage[english]{babel}

\usepackage{comment}

\usepackage[
square,numbers
]{natbib}

\hypersetup{
    unicode=false,          % non-Latin characters in Acrobat’s bookmarks
    pdftoolbar=false,        % show Acrobat’s toolbar?
    pdfmenubar=false,        % show Acrobat’s menu?
    pdffitwindow=true,     % window fit to page when opened
    pdfstartview={FitH},    % fits the width of the page to the window
    pdfkeywords={keyword1, key2, key3}, % list of keywords
    pdfnewwindow=true,      % links in new PDF window
    colorlinks=false,       % false: boxed links; true: colored links
    linkcolor=red,          % color of internal links (change box color with linkbordercolor)
    citecolor=green,        % color of links to bibliography
    filecolor=magenta,      % color of file links
    urlcolor=cyan           % color of external links
}

\usepackage{filecontents}

\begin{document}

\title{Horn: A System for Parallel Training and Regularizing of Large-Scale Neural Networks}
\maketitle

\begin{abstract}
  I introduce a new distributed system for effective training and regularizing of Large-Scale Neural Networks on distributed computing architectures. The experiments demonstrate the effectiveness of flexible model partitioning and parallelization strategies based on neuron-centric computation model and the parallel ensemble technique, with an implementation of the collective and parallel dropout neural networks training. Experiments are performed on MNIST handwritten digits classification including results, and achieve better performance than the existing base line.
\end{abstract}

\section{INTRODUCTION}

Training a large scale deep ANN architectures, such as Convolutional Neural Nets (CNNs) and Recurrent Neural Nets (RNNs), is challenging because the training process not only involves how to parallelize the training of large models but also it can be quite prone to over fitting due to large size of the network, even with large data sets. 
The current most popular regularization method is Dropout \cite{a}, usually accompany with mini-batch stochastic gradient descent (SGD). For large-scale deep learning models, there have been attempts to parallelize SGD-based training on distributed systems. Its basic concept is that each worker trains a copy of the model and combines their results synchronously, or updates through a centralized parameter server in asynchronous way. Generally, layer-wise model parallelism is used for batch processing, but this leads to a large communication overhead between host and device, or between hosts or devices.

In this paper, I introduce a new distributed system, which allows effective training and regularizing of Large-Scale Neural Networks based on flexible model partitioning and parallelization strategies with the neuron-centric abstraction model on distributed computing architectures.
The main concept of proposed approach is demonstrated with the collective and parallel dropout neural networks training. The basic idea is as follows: We do DropConnect \cite{b} and then assign the data split and different sub-models into a number of worker groups. Each group trains multiple disconnected sparse sub-models of the parent model so that each worker runs independently of each other. Then, it performs batch training to different sub-model with (random) Dropout neuron. Thus, it generates and trains more sub-models in parallel, and achieving better performance than the existing base line, and also the locality of computation and reduction of memory usage.

\section{PROGRAMMING MODEL AND BASIC CONCEPTS}

 Horn chose a neuron-centric computation model which allows intuitive programming with a clear structure to users and flexible model partitioning for an effective parallel training of large-scale neural networks.

\subsection{Neuron-centric APIs}

The user defines the computation that takes place at each neuron in each layer of the model, and the messages that should be passed during the forward and backward phases of computation. For example, we apply a set of weights to the input data and calculate an output in \textit{forward()} function like below:

\lstset{
frame=lines,
basicstyle=\small,
        literate=
               {=}{$\leftarrow{}$}{1}
                {##}{$\in{}$}{1}
               {==}{$={}$}{1},
}
\begin{adjustbox}{padding=37pt 10pt 0pt 10pt}
\begin{lstlisting}[language=C++, linewidth=11cm]
void forward(messages [i1, i2, ..., ]) {
    sum = 0
    for each w # [i1, i2, ..., ] do
        sum = sum + i.input * i.weight
    feedforward(apply(sum));
}
\end{lstlisting}
\end{adjustbox}

Then, we measure the margin of error of the output and adjust the weights accordingly to decrease the error in \textit{backward()} function:

\lstset{
frame=lines,
basicstyle=\small,
        literate=
               {=}{$\leftarrow{}$}{1}
                {##}{$\in{}$}{1}
                {@}{$\alpha{}$}{1}
                {^}{$\Delta{}$}{1}
               {==}{$={}$}{1},
}
\begin{adjustbox}{padding=37pt 10pt 0pt 10pt}
\begin{lstlisting}[language=C++, linewidth=11cm]
void backward(messages [i1, i2, ..., ]) {
    gradient = 0
    for each w # [i1, i2, ..., ] do
        gradient = gradient + i.delta() * i.weight();
        // weight collections
        w = w + ^w (@ * output * i.delta)
        // push updates to parameter server
        push(w);
    backpropagate(gradient * applyDerivative(output));
}
\end{lstlisting}
\end{adjustbox}

During backward pass, it sends updated gradients back to the parameter server by calling \textit{push()} function.

\subsection{Model Layers}
The user constructs a layered model representing the neural network architecture. The \textit{addLayer()} function adds layer to the neural network with its number of units, activation and neuron functions like below:

\lstset{
frame=lines,
basicstyle=\small,
        literate=
               {=}{$\leftarrow{}$}{1}
                {##}{$\in{}$}{1}
               {==}{$={}$}{1},
}
\begin{adjustbox}{padding=37pt 10pt 0pt 10pt}
\begin{lstlisting}[language=C++, linewidth=11cm]
nn.addLayer(512, ReLU.class, DropoutNeuron.class);
\end{lstlisting}
\end{adjustbox}

\subsection{Normalization of Neurons}

Some networks incorporate normalization of neurons. Horn allows the user to define the \textit{interlayer()} normalization function that can be applied to the computed activation outputs. This function gives the unnormalized activation of the neurons as a vector, and returns a normalized output of the neurons.

\lstset{
frame=lines,
basicstyle=\small,
        literate=
               {=}{$\leftarrow{}$}{1}
                {##}{$\in{}$}{1}
                {@}{$\alpha{}$}{1}
                {^}{$\Delta{}$}{1}
               {==}{$={}$}{1},
}
\begin{adjustbox}{padding=37pt 10pt 0pt 10pt}
\begin{lstlisting}[language=C++, linewidth=11cm]
Vector interlayer(Vector output) {
    return output.divide(output.sum());
}   
\end{lstlisting}
\end{adjustbox}

Typically, the performed normalization divides all the outputs of a layer of neurons by their sum. Softmax units are the most common example of normalized neurons.

\subsection{Model and Data Parallelism}

Horn was designed on top of Apache Hama \cite{c}, a Bulk Synchronous Parallel (BSP) \cite{d} computing engine and HDFS, a distributed file system of Apache Hadoop \cite{e}.

The BSP framework of Apache Hama is used for performing in parallel. Within single Hama BSP job, each task group works asynchronously using region barrier synchronization (data parallelism), and trains neural network model on assigned data sets (model parallelism) in BSP paradigm, by default.

\begin{figure}[htp]
\centering
\includegraphics[width=8cm]{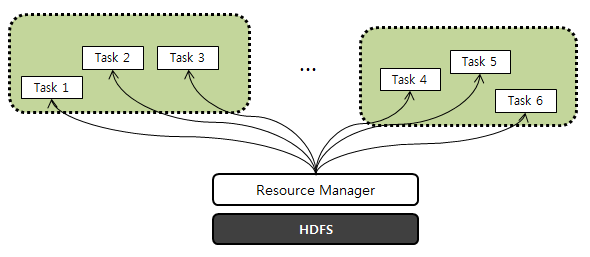}
\caption{Few tasks within the same group works in BSP paradigm. Each independent group works asynchronously.}
\end{figure}

When a neural network job is submitted to the cluster, Horn system automatically divides dataset into multiple partitions, and launches tasks (Figure 1). The task acts as a parameter server or training worker. A Zookeeper \cite{f} was used to manage the topology of parameter servers, and tasks groups and its barrier synchronizations. By configuring the cluster topology, it also allows the user to use different synchronous and asynchronous training techniques, such as AllReduce and Downpour SGD \cite{g}.

The model parallelism is done within a group. Each group divides up the neurons of network with tasks within the same group. The assigned neurons to the each task communicate directly with another by sending messages based on BSP computing model. All neurons’ functions of each layer are computed in parallel using multi-thread. During forward and backward running, the synchronization occurs in each layer of the neural network. Therefore, the locally connected learning models such as Convolutional Neural Nets (CNNs) can have more benefits and clever model partitioning is very important. 

\section{PARALLEL DROPOUT NEURAL NETWORKS}

Dropout is a technique of reducing over-fitting in neural networks by preventing complex co-adaptations on training data. It works by "dropping out" random some neuron activations in a given layer, by setting them to zero. 
Let’s take a neuron-centric Dropout Neuron where you want to use a dropout coefficient of 0.5 in layer 2 of your network. During training, the outputs of layer 2 are multiplied element-wise with a binary mask where the probability of each element of the mask being 1 is 0.5. 
\begin{center}
\[ y2 = f(z2) \cdot m2 \]
\end{center}
where f() is the activation function (e.g. Tanh, or ReLU), {$\cdot$} is an element-wise multiplication operation, z2 is the input vector of layer 2, m2 is the binary dropout mask and y2. The Dropout Neuron implemented in Horn is like below:

\lstset{
frame=lines,
basicstyle=\small,
        literate=
               {=}{$\leftarrow{}$}{1}
                {##}{$\in{}$}{1}
               {==}{$={}$}{1}
}
\begin{adjustbox}{padding=37pt 10pt 0pt 10pt}
\begin{lstlisting}[language=C++, linewidth=11cm]
void forward(messages [i1, i2, ..., ]) {
    m2 = (isTraining()) ? getBinomial(1, 0.5) : 0.5f;
    if (m2 == 0) {
        feedforward(0);
    } else {
        sum = 0
        for each w # [i1, i2, ..., ] do
            sum = sum + i.input * i.weight
        feedforward(apply(sum) * m2);
    }
}
\end{lstlisting}
\end{adjustbox}

For the training of Collective and Parallel Dropout Neural Networks, the irregular partitioning that partitions parent model into multiple sub-models (Figure 2) was used to reduce the size of model, improve the computing performance, and to get more randomness.

The main idea came from assumption that the dropout neural network can learns from many different models in parallel if they have the same input and output layers and share the weights.

\begin{figure}[htp]
\centering
\includegraphics[width=9cm]{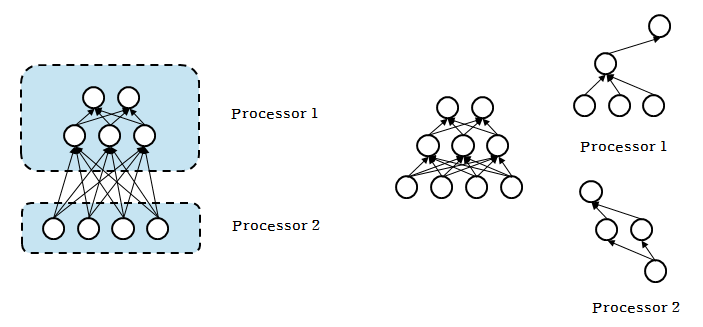}
\caption{The layer-wise partition (left) and irregularly disconnected sub-models (right).}
\end{figure}

Then each worker performs batch training to different sub-model with (random) dropout neuron. At the end of batch, each parallel set of weight updates was averaged and applied (batch averaging). The weight parameters are merged and broadcasted for the next batch in synchronous way.

\section{EXPERIMENTS}

The parallel and non-parallel dropout neural networks was tested on MNIST dataset \cite{h}, which is a large database of handwritten digits that is commonly used for training various image processing systems. In the conducted experiments, Apache Hama TRUNK version and Java 7 was used without specific JVM arguments. Both parallel and non-parallel dropout neural networks used same architecture, ReLU activation function for hidden layers, Softmax and Cross Entropy, and also same parameters: learning rate (\( \eta \) = 0.3), drop rate for input layer 0.8 and hidden layer 0.5, momentum weight (\( \alpha \) = 0.98), 100 samples per batch. 

In Non-parallel version, the batch size of single processor was 100 and the accuracy was 0.9535 at 10,000 iterations. 
In Parallel version, batch size per processor was 5 and the AllReduce training with 20 workers and single parameter server was used in this experiment. Interestingly, it trains better than non-parallel version and also gets the higher accuracy 0.9713 at 10,000 iterations. 

\begin{figure}[htp]
\centering
\includegraphics[width=9cm]{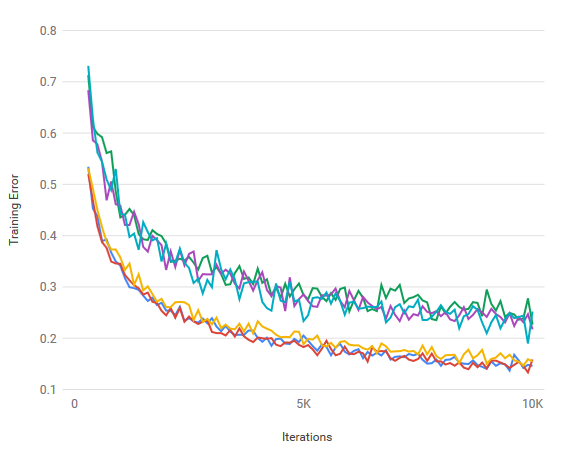}
\caption{Top is non-Parallel and bottom is Parallel. In parallel version, 20 processors were used.}
\end{figure}

Results of the experiment (Figure 3) have shown that it is possible to better train the dropout neural network using different sub models in parallel. 
Both took 30 minutes or less until 10,000 iterations. It was faster than other deep learning toolkit such as Pylearn \cite{i} and Theano \cite{j}.

\section{DISCUSSION}
\subsection{FUTURE WORKS}

The current version is only support basic Neural Network and CPU cluster. We’re now working on supporting Convolutional Neural Nets (CNNs). The known weakness is that this neuron-centric computation model can be slower than the optimized matrix multiplication, although it provides high flexibility and scalability. We also try to supporting multi-GPUs: take a neuron-centric model, and compile it to a GPU-oriented code that batches for speed. We believe that way the user will not have to think about the specifics of GPU, but instead focus on the algorithm.

\bibliographystyle{unsrt}
\bibliography{\jobname} 

\end{document}